\documentclass[aps,prb,reprint,citeautoscript,superscriptaddress,showpacs,floatfix,nofootinbib]{revtex4-2}
\usepackage[english]{babel} 
\usepackage[utf8]{inputenc} 
\usepackage{amsmath}   
\usepackage{amssymb}   
\usepackage{amsfonts}  
\usepackage{wasysym}   
\usepackage{nccmath}   
\usepackage{mathtools} 
\usepackage{bbold}     
\usepackage{graphicx}  
\usepackage{subfigure} 
\usepackage{rotating}  
\usepackage[dvipsnames]{xcolor} 

\usepackage{epsfig}
\usepackage{dcolumn}

\usepackage{endnotes}

\begin{document}
\title{Towards $\textit{ab initio}$ identification of paramagnetic substitutional carbon defects in hexagonal boron nitride acting as
	   quantum bits}

\author{Philipp Auburger}
\affiliation{Wigner Research Centre for Physics, Hungarian Academy of Sciences, P.O.\ Box 49, 1525 Budapest, Hungary}
\author{Adam Gali}
\email{gali.adam@wigner.hu}
\affiliation{Wigner Research Centre for Physics, Hungarian Academy of Sciences, P.O.\ Box 49, 1525 Budapest, Hungary}
\affiliation{Department of Atomic Physics, Budapest University of Technology and Economics, Budafoki \'{u}t 8, 1111 Budapest, Hungary}

\begin{abstract} 
 Paramagnetic substitutional carbon (C$_\text{B}$, C$_\text{N}$) defects in hexagonal boron nitride (hBN) are discussed as candidates
 for quantum bits. Their identification and suitability are approached by means of photoluminescence (PL), charge transitions, electron
 paramagnetic resonance, and optically detected magnetic resonance (ODMR) spectra. Several clear trends in these are revealed by means
 of an efficient plane wave periodic supercell \textit{ab initio} density functional theory approach. In particular, this yields 
 insight into the role of the separation between C$_\text{B}$ and C$_\text{N}$. In most of the cases the charge transition between the 
 neutral and a singly charged ground state of a defect is predicted to be experimentally accessible, since the charge transition level
 (CTL) position lies within the band gap. \textit{A posteriori} charge corrections are also discussed. A near-identification of an
 experimentally isolated single spin center as the neutral C$_\text{B}$ point defect was found via comparison of results to recently
 observed PL and ODMR spectra.
\end{abstract}
 \pacs{} 

 \maketitle 

\section{\label{intro} Introduction}
 Single-photon emitters (SPEs) in semiconductors are the most attractive fundamental building blocks for novel quantum technologies
 such as quantum computing \cite{JPhysChem.A125(6).1325,PhysRev.B97(21).214104(13),PhysRev.B102(9).099903(2),PhysRev.B97(6).064101(9),
 NatMat.19(5).534,NanoLett.16(7).4317}, quantum information distribution \cite{NatMat.20(3).321,JPhysChem.A125(6).1325,
 NatMat.19(5).534}, quantum sensing \cite{JPhysChem.A125(6).1325,NatMat.19(5).534}, quantum photonics \cite{NatMat.20(3).321,
 PhysRev.B103(11).115421(13),ApplPhysLett.115(21).212101(4)}, quantum cryptography \cite{PhysRev.B97(21).214104(13),
 PhysRev.B102(9).099903(2),PhysRev.B97(6).064101(9)}, hybrid spin-photon interfaces \cite{PhysRev.B103(11).115421(13)} and 
 spin-mechanics interfaces \cite{PhysRev.B103(11).115421(13)}.

 To create more versatile structures an embedding material with a large band gap, such as diamond
 \cite{PhysStatusSolidiA.215(22).1800318(7)}, is desirable, because atomic like defect states and levels isolated from the environment 
 can be controlled. In particular, this control over the electronic configuration also concerns charge and spin. Another issue is to 
 engineer the suitable atomic like defects for scalable quantum architecture where twodimensional (2D) materials offer a natural 
 platform to this end.

 Hence hexagonal boron nitride (hBN), an sp$^2$-bonded strongly covalent \cite{PhysRev.B103(11).115421(13)} layered van-der-Waals solid
 with indirect band gap at 5.955~eV \cite{NatPhotonics.10.262} and excellent chemical and thermal stability
 \cite{PhysRev.B97(21).214104(13),PhysRev.B102(9).099903(2)} retaining the wide band gap in its exfoliated form
 \cite{PhysRev.B103(11).115421(13)}, has recently emerged as a promising alternative \cite{NatMat-1sthBN,NatMat.20(3).321,
 JPhysChem.A125(6).1325,PhysRev.B103(11).115421(13),ApplPhysLett.115(21).212101(4),PhysRev.B97(21).214104(13),PhysRev.B102(9).099903(2),
 NatMat.19(5).534,NanoLett.16(7).4317}. It hosts a plethora of stable optically active defects with a broad emission range (1.2-5.3~eV
 \cite{PhysRev.B103(11).115421(13)}) and very high brightness, also at room temperature \cite{NatMat.19(5).534}. Moreover they exhibit
 additional favorable quantum-optical properties including narrow linewidth, high emission into the zero-phonon line (ZPL) and 
 addressability via spin-selective optical transitions \cite{NatMat.20(3).321}.  
 
 Recently, a breakthrough has been achieved by observing single spin resonance in hBN as spin-dependent fluorescence intensity, i.e., 
 continuous-wave (cw) optically detected magnetic resonance (ODMR) measurements, where the electron spin resonance (ESR) spectrum was 
 broadened by the interaction with the electron spin and the proximate nuclear spins of the host \cite{NatMat-1sthBN}. The origin of 
 the single spin centers is unknown, in stark contrast to the other ODMR center in hBN \cite{Gottscholl2020} which is the negatively 
 charged boron-vacancy defect \cite{Ivady2020,Sajid2020}. The negatively charged boron-vacancy was proposed earlier as a quantum bit
 candidate by theory~\cite{Abdi2018} which implies a strong predictive power of first principles calculations. With using the same 
 theoretical approach, this study now focuses on the recently observed single spin centers with using the advantage that the observed 
 ESR signals provide direct information about the spin density distribution of the defect in the paramagnetic ground state 
 \cite{NatMat-1sthBN}.

 Generally the atomic structure of isolated optical emitters in hBN cannot be determined directly in experiments. This is due to the 
 vast number of candidates, e.g.\ different complexes of substitutional carbon or oxygen atoms possibly with adjacent vacancies or 
 antisites. However, simple models can be developed to capture the fundamental motifs before the identification of specific defects. 
 Substitutional carbon defects are popular candidates, since they exhibit low formation energies \cite{PhysRev.B97(21).214104(13)} as 
 well as high migration barriers \cite{PhysRev.B103(11).115421(13)} and carbon interstitials are highly mobile
 \cite{JPhysChem.A125(6).1325}.
 Indeed Mendelson et al.\ \cite{NatMat.20(3).321} recently found a direct correlation between the photoluminescence of visible SPEs
 ($E_\text{ZPL}$=2.08-2.16~eV upon 532-nm illumination) and implanted carbon surveying various samples grown via metalorganic 
 vapor-phase epitaxy (MOVPE), molecular beam epitaxy (MBE) and highly oriented pyrolytic graphite (HOPG) conversion. Additionally they 
 also associated the spin readout of room-temperature ODMR performed on MOVPE samples with the density of carbon-related quantum 
 emitters which was varied from singles to ensembles. Their X-ray photospectroscopy (XPS) experiments yielded even more evidence for
 the presence of carbon point defects and complexes thereof. The neutral C$_\text{B}$C$_\text{N}$ dimer has been proposed to explain 
 the ubiquitous 4.1-eV photoluminescence \cite{ApplPhysLett.115(21).212101(4)}. Since a stepwise change of ZPL energies has been 
 reported for the single spin centers \cite{NatMat-1sthBN}, not only isolated or adjacent point defects, but also structures with 
 increased separation between C$_\text{B}$ and C$_\text{N}$ are analyzed. At a later stage the results might also be helpful for 
 target-oriented engineering of charge transition level (CTL) positions in quantum technologies. We note that, parallel to our work, 
 the optical signatures of neutral C$_\text{B}$ and C$_\text{N}$ pair and trimer configurations have been considered by first
 principles studies \cite{JPhysChem.A125(6).1325,PhysRev.B103(11).115421(13)}. In our study, beside the optical signals, we 
 systematically investigate CTLs associated with the photostability upon illumination and the ESR signals associated with ODMR spectra 
 too, including charged C$_\text{B}$ and C$_\text{N}$ pair configurations that are paramagnetic in their ground state.

 This paper is organized as follows. Section \ref{theory} elucidates the theoretical model. Section~\ref{results} discusses the results 
 for substitutional carbon (C$_\text{B}$, C$_\text{N}$) defects focusing on paramagnetic structures. It successively elaborates on
 photoluminescence (Sec.~\ref{PL-spectra}), charge transition levels (Sec.~\ref{CTL-positions}) and cw EPR/ODMR spectra
 (Sec.~\ref{cw-EPR-ODMR}). The results are summarized in Sec.~\ref{summary}. Finally, Sect.~\ref{final} consolidates the implications 
 for defect identification and quantum bit engineering, and it provides an outlook. The Supplemental Material \cite{SM} is a collection 
 of data regarding hyperfine coupling and electronic electric field gradients (EFGs) at the atomic nuclei.

\section{\label{theory}Methodology}
 The investigation is conducted within the framework of plane wave periodic supercell $\textit{ab initio}$ density functional theory
 (DFT) using the VASP package https://www.vasp.at/ \cite{PhysRev.B54(16).11169(18),PhysRev.B59(3).1758}. The systems are intralayer 
 defects and respective defect complexes consisting of one to three substitutional carbon atoms embedded in bulk hexagonal boron 
 nitride (hBN). The bulk cells are hexagonal ($8\times8\times1$) cells. All calculations are performed for the $\Gamma$ point.

 Screened hybrid DFT, the modified Heyd-Scuzeria-Ernzerhof (HSE) \cite{JChemPhys.118(18).8207,JChemPhys.124(21).219906(1)} with 
 dispersion corrections, 32\% exact exchange (instead of the original HSE06 at 25\%), screening parameter $\omega = 
 0.2/\mathring{\text{A}}$) is used to describe electronic structures and to calculate total energies. For the given supercell size it 
 yields an indirect (direct) band gap of 5.950~eV (6.439~eV), which is almost identical to the experimental value of 5.955~eV 
 \cite{NatPhotonics.10.262} (6.42~eV \cite{ApplPhysLett.109(12).122101(5)}). Ions are relaxed only in the defective molecular layer as 
 other deviations from the ideal bulk crystal geometry turned out to be negligible. It was found that it is more efficient to optimize 
 the geometry of the systems with semilocal generalized gradient approximation DFT (PBE) \cite{PhysRevLett.77(18).3865,
 PhysRevLett.78(7).1396} including dispersion corrections. Only the hybrid DFT lattice constants $a = 2.487$~\AA\ and $c = 6.459$~\AA\ 
 have to be adjusted afterwards. They are $\le2\%$ smaller than the experimental values \cite{ApplPhys.A75(3).431}. Dispersion 
 corrections (Grimme-D3 \cite{JChemPhys.132(15).154104(19),JPhysChemLett.7(12).2197(7)} with Becke-Johnson (BJ) damping 
 \cite{JComputChem.32(7).1456}) are indispensable for an adequate treatment of the interlayer interaction and hence the determination 
 of the distance ($c/2$) between adjacent layers.

 The photoluminescence (PL) spectra are determined after the additionally required data have been obtained from two separate 
 calculations \cite{NewJPhys.16(7).073026(23)}. The total energy and geometry of the excited state are obtained from a $\Delta$SCF 
 calculation \cite{Gali2009}. We note that the $\Delta$SCF method inherently contains the electron-hole interaction which is strong in hBN~\cite{Attaccalite2011}. For singlets the ZPL energy is corrected as given in Ref.~\onlinecite{ApplPhysLett.115(21).212101(4)}. 
 The ground state phonon modes are calculated on the aforementioned semilocal gradient DFT level, which is considered a reliable 
 approximation drastically reducing the computational effort \cite{Ivady2020,JPhysChem.A125(6).1325,PhysRev.B103(11).115421(13)}. Again 
 only the ions in the defective molecular layer are allowed to move. Furthermore the cw ODMR defect line broadening can also be 
 accessed, since the hyperfine coupling for the most abundant nuclear spin active isotopes (${}^{11}$B, ${}^{14}$N, ${}^{13}$C) is 
 analyzed with the aforementioned screened hybrid DFT \cite{PhysRev.B62(10).6158, PhysRev.B88(7).075202(7)}. The cw ODMR spectra are 
 calculated based on this data employing the MATLAB toolbox EasySpin \cite{PhD-thesis-Stoll-2003, JMagnReson.178(1).42, 
 STOLL-2015-121}. EasySpin can also take the nuclear Zeeman and quadrupole interaction into account. The electronic electric field 
 gradients (EFGs) at the positions of the (quadrupolar) atomic nuclei ($I\ge1$) required for the latter are calculated with VASP 
 \cite{PhysRevB.57(23).14690}. The EFG matrix $V$ is converted into the quadrupole matrix $Q$ according to 
 \begin{equation} \label{V-to-Q}
  Q = \frac{eq_n}{2I(2I-1)h}V \text{,}
 \end{equation}
 where $q_n$ and $I$ are the nuclear quadrupole moment and nuclear spin quantum number, respectively \cite{eMagRes.6(4).495}, and $h$ 
 is the Planck-constant. This leads to $Q$(MHz)=0.01636$V$(V/\AA$^2$) for ${}^{11}$B and $Q$(MHz)=0.02471$V$(V/\AA$^2$) for ${}^{14}$N. 
 Carbon atom has only non-quadrupolar isotopes.

 The plane wave energy cutoff is 500~eV throughout. The energies are considered converged for differences of less than 0.01~meV. In the
 semilocal generalized gradient approximation DFT calculations the Hellmann-Feynman force components within the defective molecular 
 layer are reduced to less than 5~meV/$\mathring{\text{A}}$. The PL spectra were calculated with a Gaussian smearing of 5~meV.

 The total energy of the charged systems, and hence the position of the charge transition levels (CTLs), is affected by a spurious 
 contribution due to the interaction between the charge and its periodic images. Hence \textit{a posteriori} charge corrections 
 according to Freysoldt et al.\ \cite{PhysStatSol.B248(5).1067, PhysRevLett.102(1).016402(4)} are applied for the charged bulk defects.

\section{\label{results} Results}
\begin{figure}[h]
  \includegraphics[width=\linewidth]{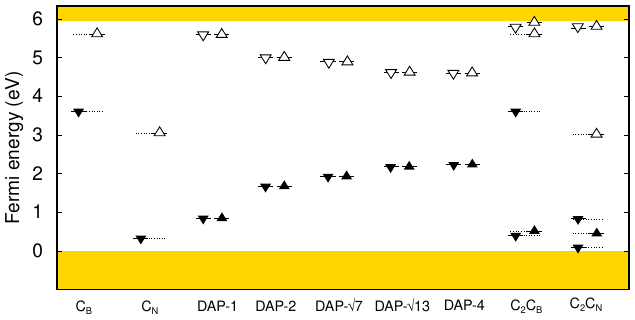}
  \caption{\small\label{fig:KSlevels} Calculated Kohn-Sham defect levels for neutral carbon defects in the fundamental band gap of hBN. DAP refers to the C$_\text{B}$C$_\text{N}$-DAP donor acceptor pair defects. The spinpolarized calculation introduces a gap between the occupied (filled triangle) and unoccupied (empty triangle) defect levels.}
 \end{figure}

 Chejanovsky et al.\ \cite{NatMat-1sthBN} recorded the PL spectrum of an isolated optical emitter (cf.\ Fig.\
 \ref{PL-candidates-PSB-compared-to-D1}) with $E_\text{ZPL}$=1.71~eV (cf.\ Tab.~\ref{ZPL-energies}) upon illumination by 633-nm 
 (1.96-eV) laser. Its ground state has to be paramagnetic with non-integer spin as concluded based on ODMR (c.f.\ 
 Fig.~\ref{D1-ODMR-compared-to-candidates}) and photokinetics. Furthermore, the electronic spin density is essentially out-of-plane and 
 $\pi$-like as indicated by angular-dependence and magnitude of the hyperfine coupling. We concluded that paramagnetic substitutional 
 carbon defects with $S=1/2$ spin states can satisfy these criteria (see Fig.~\ref{fig:KSlevels} for electronic structure). Since the abundance of $I=1/2$ ${}^{13}$C is low no hyperfine 
 splitting due to the carbon impurity is expected for the vast majority of the spin centers.
 
\begin{figure}[h]
  \includegraphics[width=\linewidth]{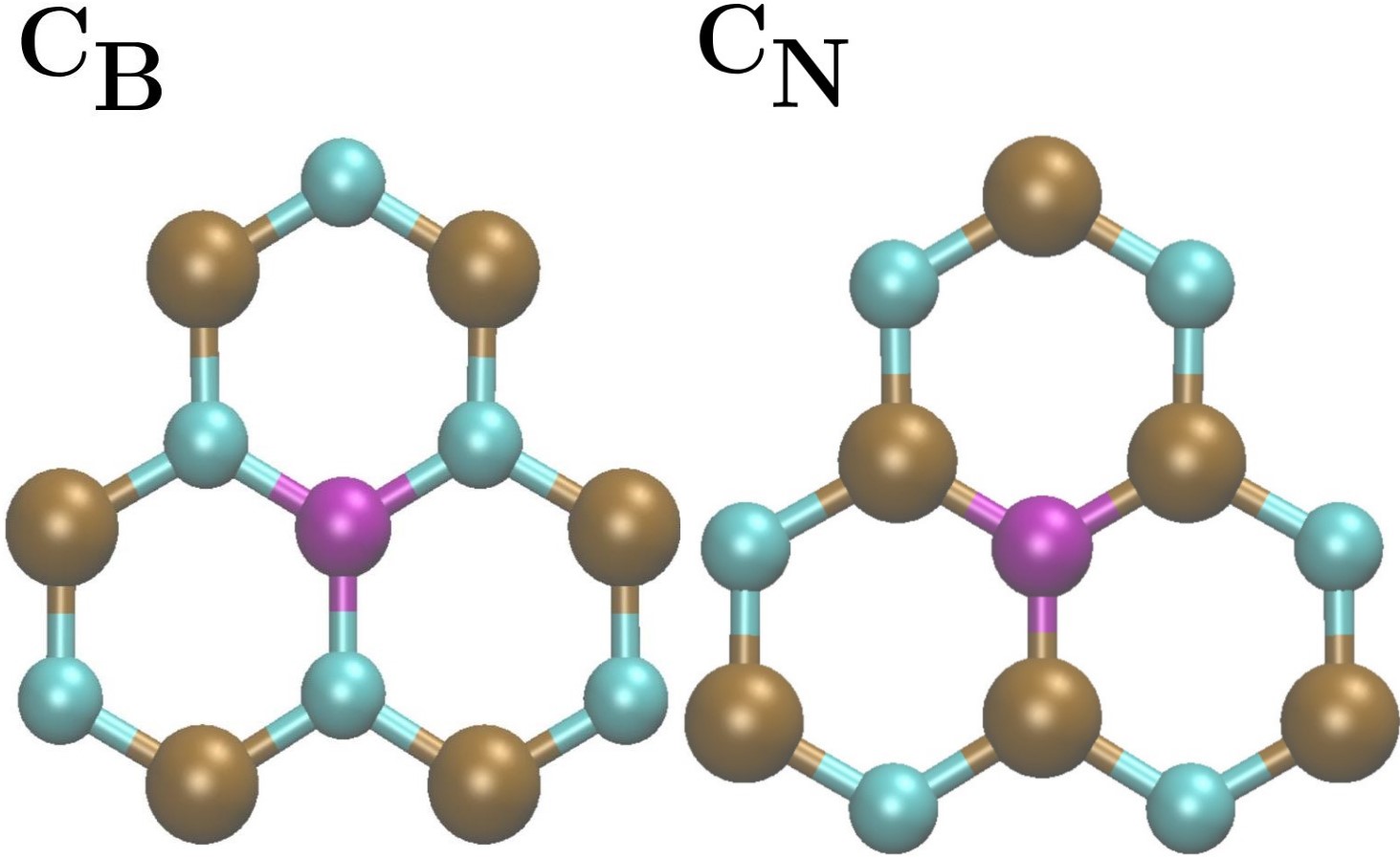}
  \caption{\small\label{CB-and-CN} Substitutional carbon defects in hBN (part 1, neutral paramagnetic, from left to right): 
  	       C$_\text{B}$, C$_\text{N}$; (B in ochre, N in cyan, C in purple)}
 \end{figure}

\begin{figure}[h]
  \includegraphics[width=\linewidth]{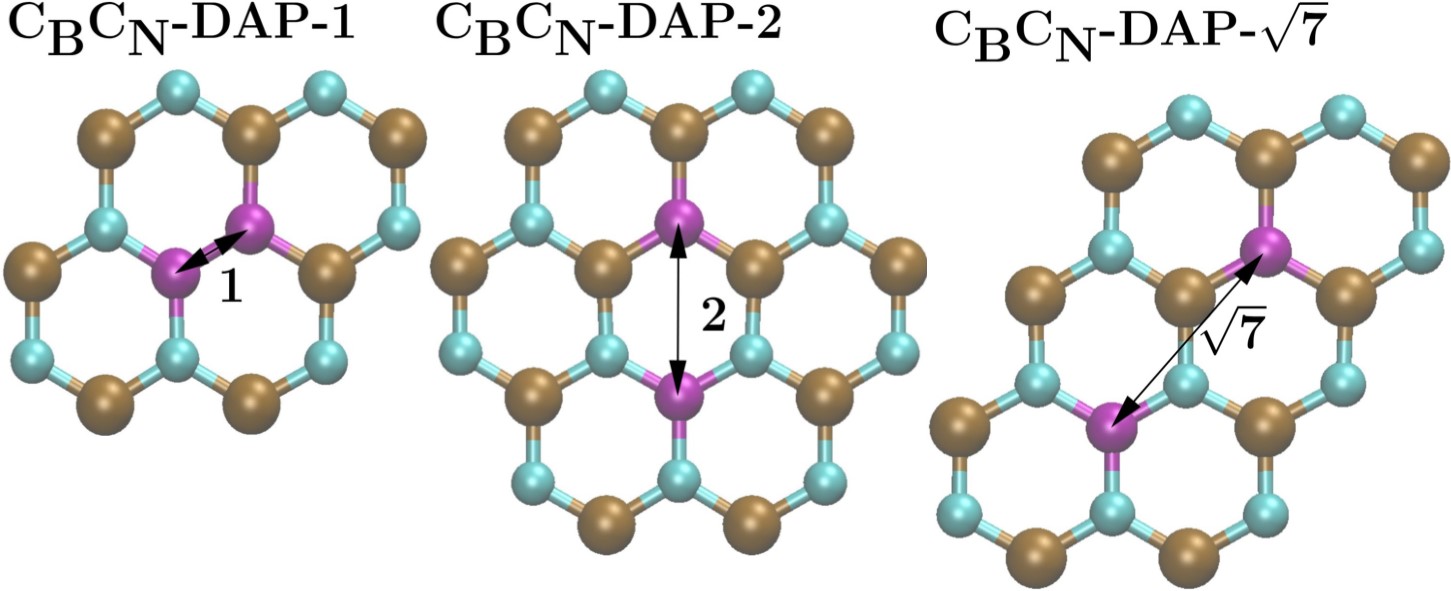}
  \caption{\small\label{CBCN-DAP-1-and-2-and-sqrt7-half-size} Substitutional carbon defects in hBN (part 2, singly charged 
  	       paramagnetic, from left to right): C$_\text{B}$C$_\text{N}$-DAP-1 (dimer), C$_\text{B}$C$_\text{N}$-DAP-2, 
  	       C$_\text{B}$C$_\text{N}$-DAP-$\sqrt{7}$; (B in ochre, N in cyan, C in purple, cf.\ text for DAPs)}
 \end{figure}

\begin{figure}[h]
  \includegraphics[width=\linewidth]{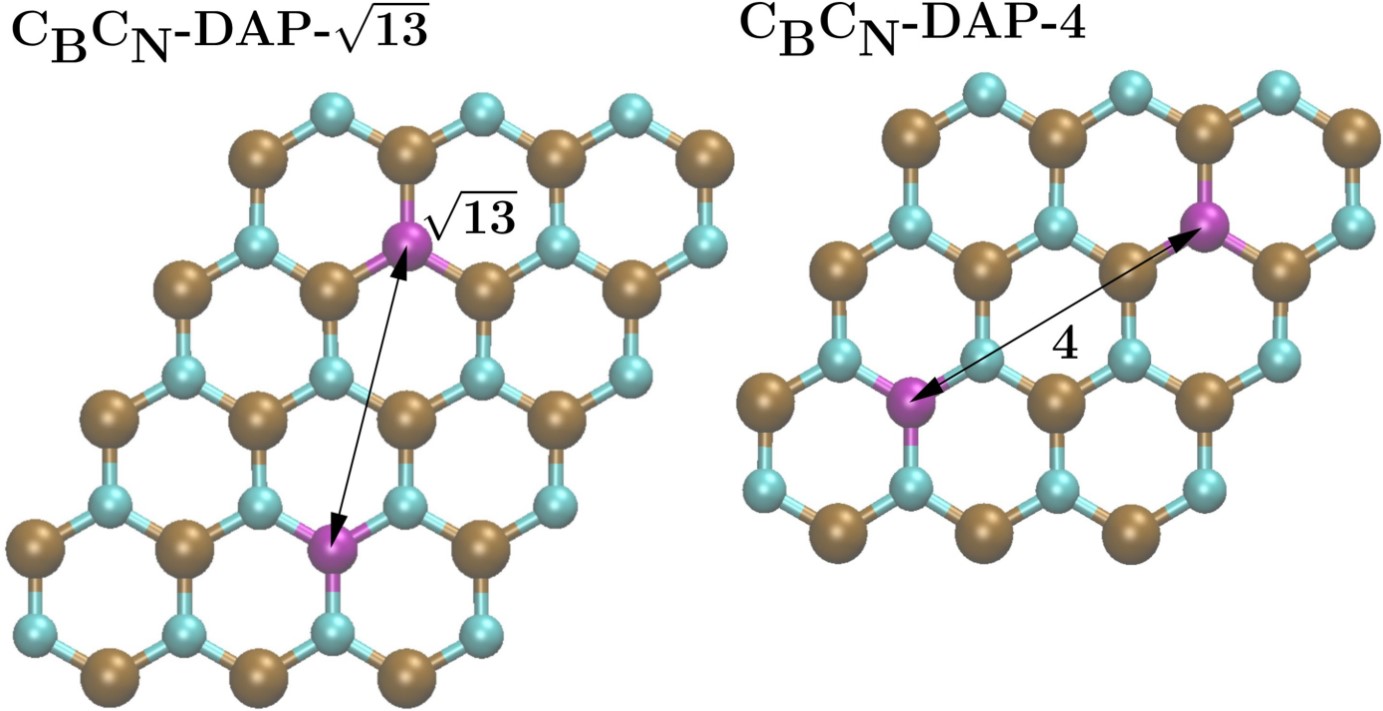}
  \caption{\small\label{CBCN-DAP-sqrt13-and-4-half-size} Substitutional carbon defects in hBN (part 3, singly charged paramagnetic, 
  	       from left to right): C$_\text{B}$C$_\text{N}$-DAP-$\sqrt{13}$, C$_\text{B}$C$_\text{N}$-DAP-4; (B in ochre, N in cyan, C in purple, cf.\ text for DAPs)}
 \end{figure}

\begin{figure}[h]
  \includegraphics[width=\linewidth]{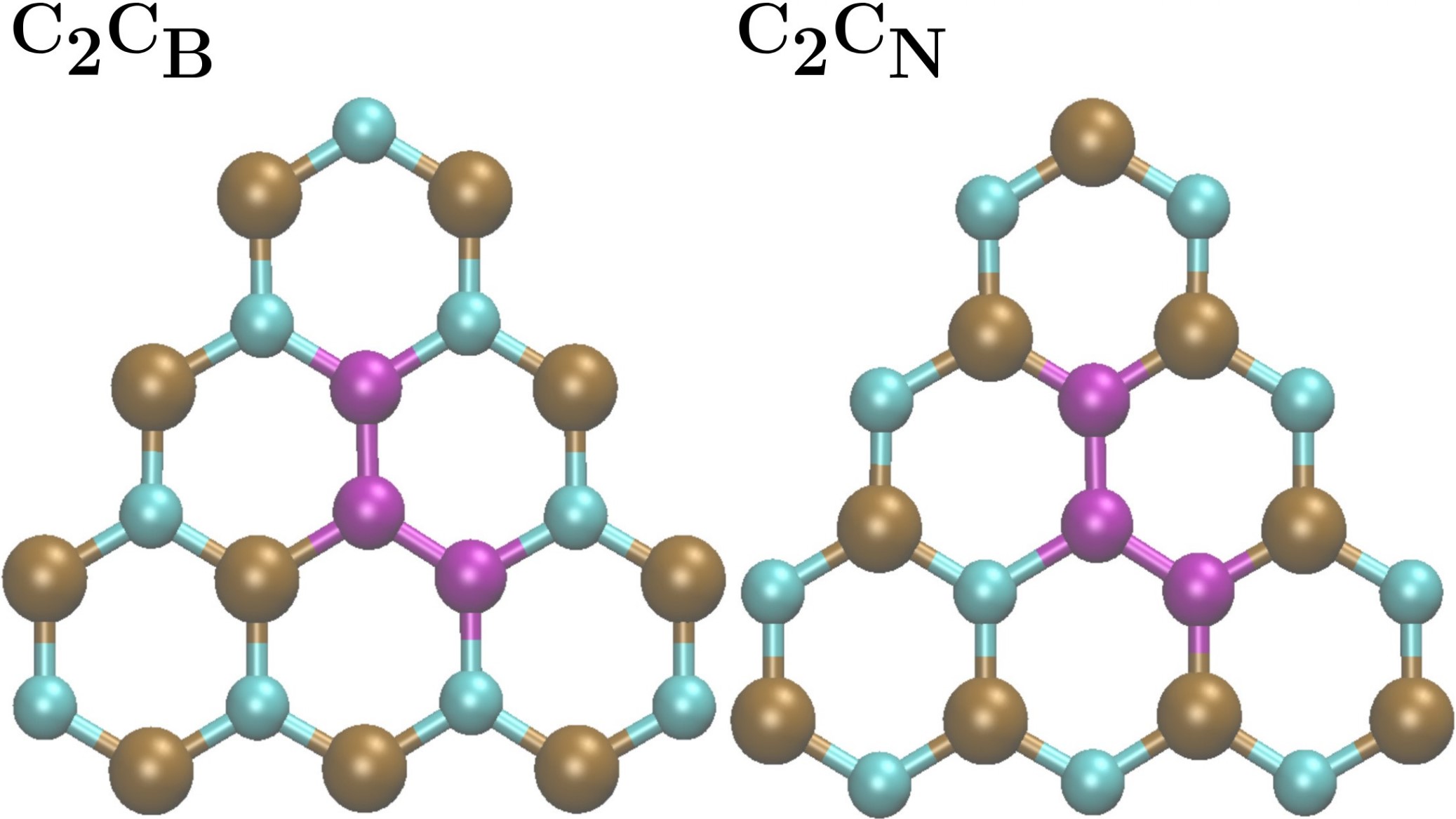}
  \caption{\small\label{C2CB-and-C2CN} Substitutional carbon defects in hBN (part 4, neutral paramagnetic, from left to right):   
  	       C$_\text{2}$C$_\text{B}$-trimer, C$_\text{2}$C$_\text{N}$-trimer; (B in ochre, N in cyan, C in purple)}
 \end{figure}

 With up to three adjacent carbon atoms there are six basic ways to construct a paramagnetic defect (complex) exhibiting paramagnetic 
 $S=1/2$ states: two neutral isolated point defects (C$_\text{B}$, C$_\text{N}$), two singly charged dimers 
 (C$_\text{B}$C$_\text{N}$(+), C$_\text{B}$C$_\text{N}$(-)) and two neutral trimers (C$_2$C$_\text{B}$, C$_2$C$_\text{N}$) (cf.\ 
 Figs.~\ref{CB-and-CN}, \ref{CBCN-DAP-1-and-2-and-sqrt7-half-size}, and \ref{C2CB-and-C2CN}, respectively). Starting from the 
 C$_\text{B}$C$_\text{N}$ dimer structures with larger separation between the point defects are also investigated. These are called 
 ``donor-acceptor pairs'' (DAPs), because for the neutral case an electron transfer resulting in a non-paramagnetic structure 
 (C$_\text{B}^+$ and C$_\text{N}^-$) is energetically favorable (see the text below as well as Secs.~\ref{PL-spectra} and \ref{CTL-positions}). The DAPs are 
 labeled according to their separation in the ideal lattice with ``1'' being the dimer. The five smallest separations are 1, 2, 
 $\sqrt{7}$, $\sqrt{13}$, 4 (cf.\ Figs.~\ref{CBCN-DAP-1-and-2-and-sqrt7-half-size}, \ref{CBCN-DAP-sqrt13-and-4-half-size}).

We plot the in-gap Kohn-Sham levels of the neutral defects in Fig.~\ref{fig:KSlevels}. The isolated neutral C$_\text{B}$ and C$_\text{N}$ defects introduce a single defect level into the fundamental band gap occupied by a single electron. The occupied level of C$_\text{B}$ lies at higher energy than the empty level of C$_\text{N}$, thus the electron transfer occurs from C$_\text{B}$ towards C$_\text{N}$ when the defects are relatively close to each forming DAP complexes as noted above. This reflects in the calculated defect levels of DAP defects which show a completely filled low-energy level (localized around C$_\text{N}$) and an empty high-energy level (localized around C$_\text{B}$). By combining three carbon defects, i.e., the considered trimer complexes in our study, the electronic structure shows the characteristic of the constituent defects. In C$_\text{2}$C$_\text{N}$-trimer, two low-energy defect levels appear localized around C$_\text{N}$ atoms and a high-energy level localized around C$_\text{B}$ atom where the two low-energy defect levels are occupied by three electrons. In C$_\text{2}$C$_\text{B}$-trimer, rather one low-energy defect level and two high-energy defect levels appear in the gap occupied by three electrons. 

The calculated binding energies are listed in Table~\ref{tab:binding_energies}. The donor-acceptor pairs are stabilized through charge transfer from C$_\text{B}$ donor to C$_\text{N}$ acceptor which involves Coulomb-interaction. As the attractive Coulomb-interaction increases with shorter distances the binding between C$_\text{B}$ and C$_\text{N}$ is stronger for pairs with shorter distances. The calculated binding energies clearly show this trend. We note that this looks to hold for the immediate C-C dimer defect but a C-C bond is created in that defect, and it cannot be described by the simple donor-acceptor pair model. Since the isolated carbon impurities introduce defect wavefunctions localized inside the host sheet and interlayer diffusion of carriers is hindered in hBN, complexes of carbon substitutional defects within a single sheet of hBN layer are considered in which we expect direct electron transfer between the constituting carbon defects. Certainly, different configurations of DAP and trimer complexes may occur in each hBN layer in multilayered hBN but we do not expect direct interaction between carbon defects residing in different layers of hBN.
\begin{table}
\caption{\small\label{tab:binding_energies}The calculated binding energy of neutral carbon complexes within a layer of hBN. The negative binding energy refers to the favor of complex formation.}
\begin{ruledtabular}
\begin{tabular}{lr}
Defect & Binding energy (eV) \\
\hline
C$_\text{B}$C$_\text{N}$-DAPs & \\
1               &  -3.93 \\
2               & -2.04 \\
$\sqrt{7}$  & -1.90 \\
$\sqrt{13}$ & -1.52 \\
4                & -1.49  \\
trimers & \\
C$_2$C$_\text{B}$  & -5.31 \\
C$_2$C$_\text{N}$  & -5.28 \\
\end{tabular}
\end{ruledtabular}
\end{table}     


\subsection{\label{PL-spectra} Photoluminescence}
 %
 The ZPL energy $E_\text{ZPL}$ is the first criterion for potentially identifying an experimentally isolated emitter. It has been 
 calculated for all neutral systems and additionally also for the singly charged DAPs (cf.\ Tab.~\ref{ZPL-energies}). The theoretical 
 values range from 0.44~eV to 4.13~eV. $E_\text{ZPL}$ depends on the electronic structure of the defect (complex), in particular its
 states inside the band gap, and its ionic relaxation triggered via deexcitation resp.\ emission.

 The first optical excitation of a neutral point defect either transfers the unpaired electron from the band gap to the conduction band
 minimum (C$_\text{B}$, $E_\text{ZPL}$=1.695~eV, cf.\ Fig.~\ref{ZPL-energies}) or instead transfers another electron from the valence 
 band maximum into its orbital (C$_\text{N}$, $E_\text{ZPL}$=2.468~eV, cf.\ Fig.~\ref{ZPL-energies}). The former electron transfer is 
 within the majority spin channel, while it is the minority spin channel for the latter. This behavior was obtained earlier by Jara et 
 al.\ \cite{JPhysChem.A125(6).1325}, but they gave no values for $E_\text{ZPL}$.

 Although the neutral point defects (C$_\text{B}$, C$_\text{N}$) are paramagnetic, their interaction leads to non-paramagnetic 
 structures (C$_\text{B}$C$_\text{N}$-DAPs with ``charge structure'' C$_\text{B}^+$-C$_\text{N}^-$), because transferring the unpaired 
 electron from C$_\text{B}$ to C$_\text{N}$ is energetically favorable. The CTLs C$_\text{B}$(+1$\mid$0) and C$_\text{N}$(0$\mid$-1) 
 (cf.\ Tab.~\ref{CTL-table}) explain this for the limiting case of large separations. All numerical calculations at finite separations 
 yielded the same behavior. Each C$_\text{B}$C$_\text{N}$-DAP exhibits two orbitals inside the band gap. The lower one is fully 
 occupied and localized at C$_\text{N}$, while the higher one is empty and localized at C$_\text{B}$. The energy difference between 
 them diminishes with increasing separation corroborating the results of Linder\"alv et al.\ \cite{PhysRev.B103(11).115421(13)}.

 $E_\text{ZPL}$ of the neutral C$_\text{B}$C$_\text{N}$-DAPs diminishes with separation (cf.\ Tab.~\ref{ZPL-energies}) as expected in 
 view of the previous findings. The neutral C$_\text{B}$C$_\text{N}$ dimer has been proposed to explain the ubiquitous 4.1-eV
 photoluminescence \cite{ApplPhysLett.115(21).212101(4)}. The calculated $E_\text{ZPL}$=4.13~eV is consistent with that and agrees well 
 with the hybrid DFT (HSE) results of Mackoit-Sinkevi\v{c}ien\.{e} et al.\ (4.31~eV in 
 Ref.~\onlinecite{ApplPhysLett.115(21).212101(4)}), who also employed a bulk model, and Jara et al.\ (4.12~eV in 
 Ref.~\onlinecite{JPhysChem.A125(6).1325}), who employed a single layer model instead. On the contrary the underlying gradient DFT 
 (PBE) is not sufficient to reproduce this result as found by Linder\"alv et al.\ (3.34~eV in 
 Ref.~\onlinecite{PhysRev.B103(11).115421(13)}).

 The ``charge structure'' of the neutral C$_\text{B}$C$_\text{N}$-DAPs is C$_\text{B}^+$-C$_\text{N}^-$. Hence a singly charged DAP can 
 be created either by removing the excess electron from C$_\text{N}^-$ (positive DAP, C$_\text{B}^+$-C$_\text{N}^0$) or by adding an 
 electron to C$_\text{B}^+$ (negative DAP, C$_\text{B}^0$-C$_\text{N}^-$). In both cases the unpaired electron is localized at one 
 point defect and the excess charge at the other one.

 $E_\text{ZPL}$ of both singly charged C$_\text{B}$C$_\text{N}$-DAPs increases with separation (cf.\ Tab.~\ref{ZPL-energies}). It seems 
 to approach a value equal, or at least similar, to that of the respective isolated neutral (paramagnetic) point defect. This 
 convergence behavior had to be expected, because it turned out that the lowest optical excitation is linked to the same electronic 
 transition as for the isolated neutral point defect. Hence the influence of the charged (non-paramagnetic) point defect vanishes with 
 separation. The aforementioned Jara et al.\ \cite{JPhysChem.A125(6).1325} also found that $E_\text{ZPL}$ of the singly charged dimers
 is too low for the visible range.

 Placing a point defect (C$_\text{B}$ or C$_\text{N}$) adjacent to the dimer (C$_\text{B}$C$_\text{N}$, abbreviated as C$_2$) creates a 
 basic trimer (C$_2$C$_\text{B}$ or C$_2$C$_\text{N}$ with the desired $S=1/2$ paramagnetic ground state, cf.\ 
 Fig.~\ref{C2CB-and-C2CN}). $E_\text{ZPL}$ of the neutral basic trimers (1.36~eV for C$_2$C$_\text{B}$, 1.62~eV for C$_2$C$_\text{N}$) 
 is lower than that of their neutral ``building blocks'' point defect and dimer (cf.\ Tab.~\ref{ZPL-energies}). This is essentially due
 to the (overlap of) additional states inside the band. Jara et al.\ \cite{JPhysChem.A125(6).1325} obtained a similar value for 
 C$_2$C$_\text{N}$ (1.62~eV), but a clearly higher one for C$_2$C$_\text{B}$ (1.65~eV). The origin of the discrepancy in the latter is 
 unclear.

\begin{figure}[h]
  \includegraphics[width=\linewidth]{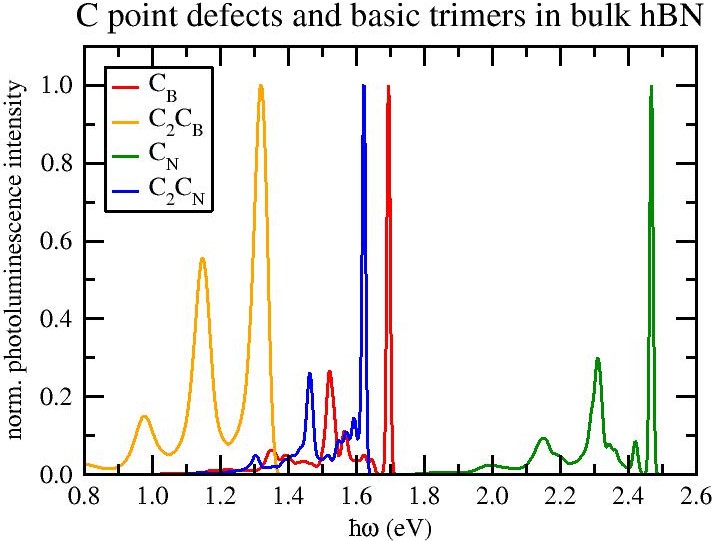}
  \caption{\small\label{PL-point-defects-and-basic-trimers-neutral} Calculated photoluminescence (PL) spectrum for the neutral point  
  	       defects (C$_\text{B}$ and C$_\text{N}$, cf.\ Fig.~\ref{CB-and-CN}) and basic trimers (C$_2$C$_\text{B}$ and 
  	       C$_2$C$_\text{N}$, cf.\ Fig.~\ref{C2CB-and-C2CN}) in bulk hBN}
 \end{figure}

\begin{figure}[h]
  \includegraphics[width=\linewidth]{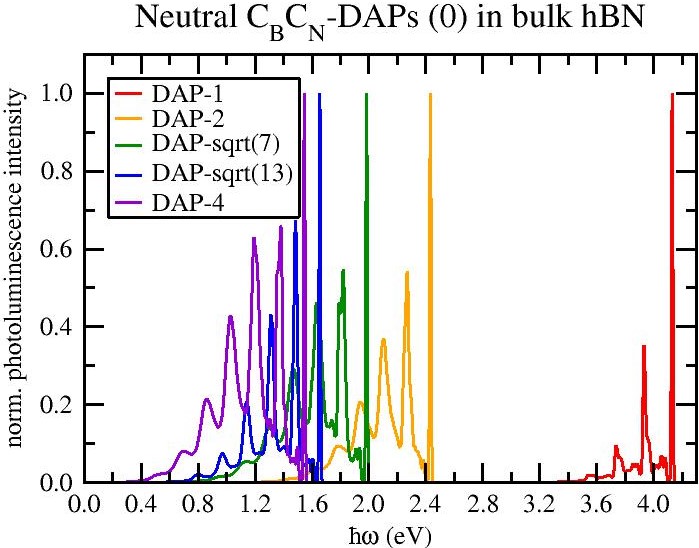}
  \caption{\small\label{PL-CBCN-DAP-neutral} Calculated photoluminescence (PL) spectrum for the five closest neutral 
  	       DAPs (cf.\ Figs.~\ref{CBCN-DAP-1-and-2-and-sqrt7-half-size} and \ref{CBCN-DAP-sqrt13-and-4-half-size}) in bulk hBN}
 \end{figure}

\begin{figure}[h]
  \includegraphics[width=\linewidth]{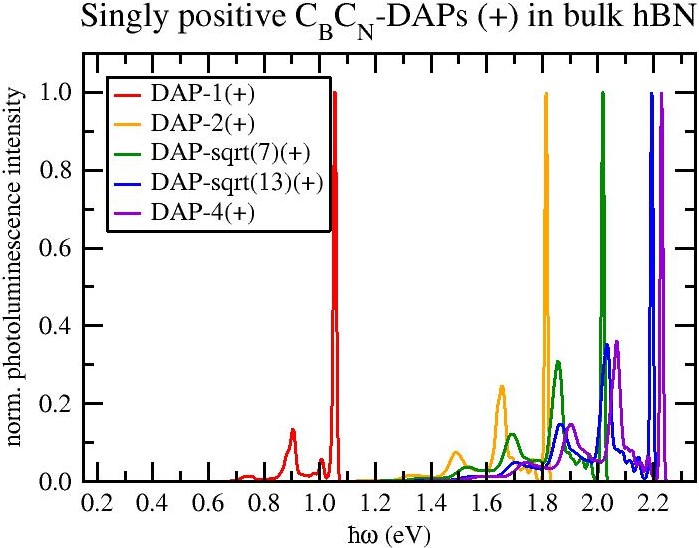}
  \caption{\small\label{PL-CBCN-DAP-positive} Calculated photoluminescence (PL) spectrum for the five closest singly positive 
  	       DAPs (cf.\ Figs.~\ref{CBCN-DAP-1-and-2-and-sqrt7-half-size} and \ref{CBCN-DAP-sqrt13-and-4-half-size}) in bulk hBN}
 \end{figure}

\begin{figure}[h]
  \includegraphics[width=\linewidth]{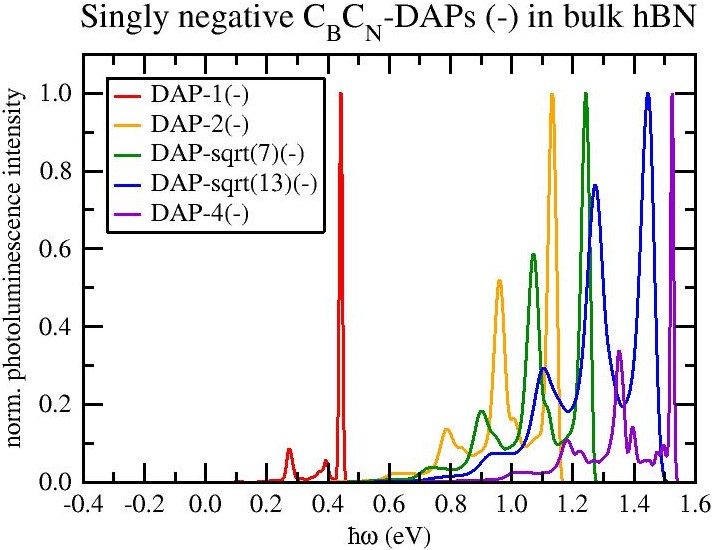}
  \caption{\small\label{PL-CBCN-DAP-negative} Calculated photoluminescence (PL) spectrum for the five closest singly negative 
   	       DAPs (cf.\ Fig.~\ref{CBCN-DAP-1-and-2-and-sqrt7-half-size} and \ref{CBCN-DAP-sqrt13-and-4-half-size}) in bulk hBN}
 \end{figure}

 Beside the characteristic ZPL emission line  (cf.\ 
 Figs.~\ref{PL-point-defects-and-basic-trimers-neutral}, \ref{PL-CBCN-DAP-neutral}, \ref{PL-CBCN-DAP-positive}, and 
 \ref{PL-CBCN-DAP-negative}), the most intense feature in the phonon sideband (PSB) always exhibited an energy detuning of 0.15-0.17~eV 
 with respect to ZPL peak, except for the neutral dimer (0.20~eV). The intensity of its replica decreases with the number of quanta. 
 Linder\"alv et al.\ \cite{PhysRev.B103(11).115421(13)} obtained similar results including a mostly somewhat larger PSB energy 
 detuning of 0.18-0.20~eV for the neutral points defects and DAPs. According to them, these phonons are slightly distorted 
 high-frequency in-plane modes of the ideal lattice, which would be at 0.15~eV, with the only exception (local mode) being the dimer 
 (C-C bond). The present results are corroborated by Jara et al.\ \cite{JPhysChem.A125(6).1325}, who report 0.20~eV resp.\ 0.16~eV for 
 neutral C$_\text{B}$C$_\text{N}$ resp.\ C$_2$C$_\text{N}$ and ascribe both to a local vibration mode. They also mention values in the 
 same range for other substitutional carbon defects. Experimental PL spectra with similar features have been recorded by various groups 
 \cite{NatMat-1sthBN,NatMat.20(3).321, PhysRev.B78(15).155204(8), NanoLett.16(7).4317}. E.g.\ Mendelson et al.\ \cite{NatMat.20(3).321} 
 recently observed several emitters with $E_\text{ZPL}$=2.08-2.16~eV and 0.16-0.20~eV PSB energy detuning upon illumination by 532-nm 
 (2.33-eV) laser. The characteristics of the ODMR spin readout are in line with those found in a previous report \cite{NatMat-1sthBN} 
 giving evidences for their connection to carbon-related defects although using a longer excitation wavelength (at least 594~nm or 
 2.09~eV) insufficient for exciting all of these emitters. Hence the singly positive C$_\text{B}$C$_\text{N}$-DAPs at larger
 separations (at least $\sqrt{7}$, cf.\ Tab.~\ref{ZPL-energies}) may explain the observations if the defect ground state spin is 
 $S=1/2$.

\begin{figure}[h]
  \includegraphics[width=\linewidth]{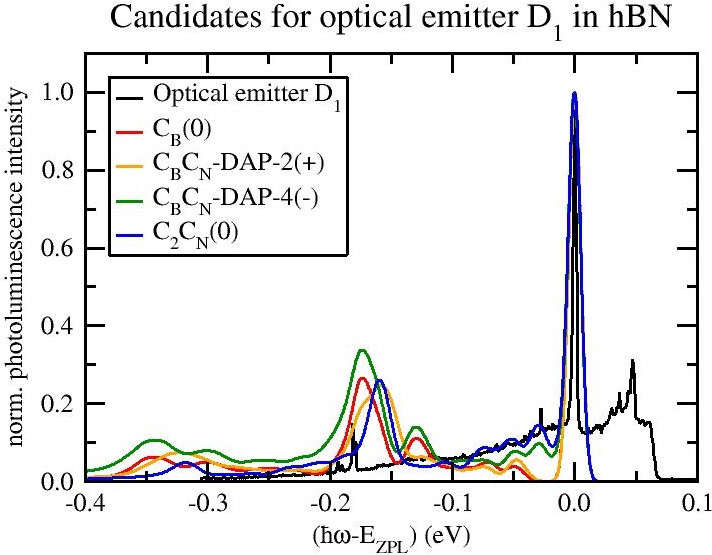}
  \caption{\small\label{PL-candidates-PSB-compared-to-D1} Comparing calculated phonon sidebands (PSBs) of candidates (cf.\
  	       Figs.~\ref{CB-and-CN}-\ref{C2CB-and-C2CN}) for identification of optical emitter D$_1$ experimentally isolated by 
  	       Chejanovsky et al.\ \cite{NatMat-1sthBN} in hBN}
 \end{figure}

 Fig.~\ref{PL-candidates-PSB-compared-to-D1} compares the PSBs of optical emitter D$_1$, which was experimentally isolated by
 Chejanovsky et al.\ \cite{NatMat-1sthBN}, to those of the candidates for its identification. D$_1$ has $E_\text{ZPL}$=1.71~eV with a
 very similar PSB to the calculated ones and is paramagnetic with half-integer spin.  We note that the other reported single photon
 emitters with longer wavelength ZPLs than that of 1.71-eV D$_1$ center in Ref.~\onlinecite{NatMat-1sthBN} have different PL lineshape,
 so the characteristic PSBs are distinct from that of D$_1$. On the other hand, all the considered carbon substitutional defects have
 very similar PSB features. Thus our paper now focuses on the identification of D$_1$ center and it is beyond the scope to identify the
 other single photon emitters.
 
 An error of 0.2~eV was allowed to take any kind of inaccuracy into account when the D$_1$ ZPL energy is compared to the calculated ZPL 
 energies.  Hence the present study provides four candidates: C$_\text{B}$(0), C$_\text{B}$C$_\text{N}$-DAP-2(+), 
 C$_\text{B}$C$_\text{N}$-DAP-4(-), C$_2$C$_\text{N}$(0) (cf.\ Tab.~\ref{ZPL-energies}). We note that we often anticipate an accuracy 
 of about 0.1~eV for the calculated ZPLs. Here we expect a larger error for some of the considered defects as excitation to band edges 
 is involved in the excited state which is a subject of finite size error for the neutral defects. Tests on  C$_\text{B}$ defect shows 
 that the calculated $E_\text{ZPL}$ increases by about 0.12~eV going from $8\times8\times1$ to $11\times11\times1$ supercell. 
 Nevertheless, we provide the calculated ZPL energies within meV throughout the paper which is technically convergent in terms of the 
 given supercell size ($8\times8\times1$) and DFT functional.

 \begin{table}[h]
  \caption{\small\label{ZPL-energies} Zero-phonon-line energy ($E_\text{ZPL}$) of substitutional carbon defects in bulk hBN obtained from screened hybrid DFT (cf.\ text and Figs.~\ref{CB-and-CN}-\ref{C2CB-and-C2CN}) and optical emitter D$_1$ (see Ref.~\onlinecite{NatMat-1sthBN}). For C$_\text{B}$C$_\text{N}$-DAPs  we provide ZPL energies at various charge states.} 
  \begin{ruledtabular}
   \begin{tabular}{l|ccc}
   	Defect & \multicolumn{3}{c}{$E_\text{ZPL}$ (eV)} \\
   	\hline
   	C$_\text{B}$($0$) & \multicolumn{3}{c}{1.695} \\
   	C$_\text{N}$($0$) & \multicolumn{3}{c}{2.468} \\
   	C$_\text{B}$C$_\text{N}$-DAPs & ($+$) & ($0$) & ($-$) \\
   	   	       4 & 2.229 & 1.547 & 1.525 \\
   	 $\sqrt{13}$ & 2.194 & 1.655 & 1.446 \\
   	  $\sqrt{7}$ & 2.018 & 1.984 & 1.243 \\
   	           2 & 1.814 & 2.432 & 1.133 \\
   	   1 (dimer) & 1.055 & 4.131 & 0.442 \\
   	Trimers & \multicolumn{2}{c}{} & \\
   	C$_2$C$_\text{B}$($0$) & \multicolumn{3}{c}{1.360} \\
   	C$_2$C$_\text{N}$($0$) & \multicolumn{3}{c}{1.623} \\
   	Emitter D$_1$ & \multicolumn{3}{c}{1.71\textcolor{white}{0}} \\ 
   \end{tabular}
  \end{ruledtabular}
 \end{table}

\subsection{\label{CTL-positions} Charge transition levels}
 To modify the charge state of a specific defect $X$ it has to be understood how it depends on the Fermi level $E_\text{F}$. The 
 transition between two charge states happens at the CTL position $E_\text{F}^{CT}$ where they are equally favorable. The defect charge 
 $q$ can be modified accordingly if this transition is between the two most stable states and $E_\text{F}^{CT}$ lies within  the band 
 gap.

 The CTLs between neutral and singly charged ground states with respect to valence band maximum ($\epsilon_\text{VBM}$) are calculated 
 as
 \begin{equation} 
  \label{CTL-calculation}
  \begin{gathered}
   E_{\text{F}(X)}^{\text{CT}(+1\mid0)}+\epsilon_\text{VBM} = E_{0(X)}^{0} - E_{0(X)}^{+1} - \Delta_{+1}(X) \\
   E_{\text{F}(X)}^{\text{CT}(0\mid-1)}+\epsilon_\text{VBM} = E_{0(X)}^{-1} - E_{0(X)}^{0} + \Delta_{-1}(X)
  \end{gathered}
 \end{equation}
 where $E_{0(X)}^{q}$ is the total energy of the defect $X$ with charge $q$ and $\Delta_{q}(X)$ is the charge correction. We note that 
 we believe that the threedimensional model corresponds to the experimental conditions where the defects were found in multilayer hBN 
 on a substrate which introduce a charge screening.

\begin{figure}[h]
  \includegraphics[width=\linewidth]{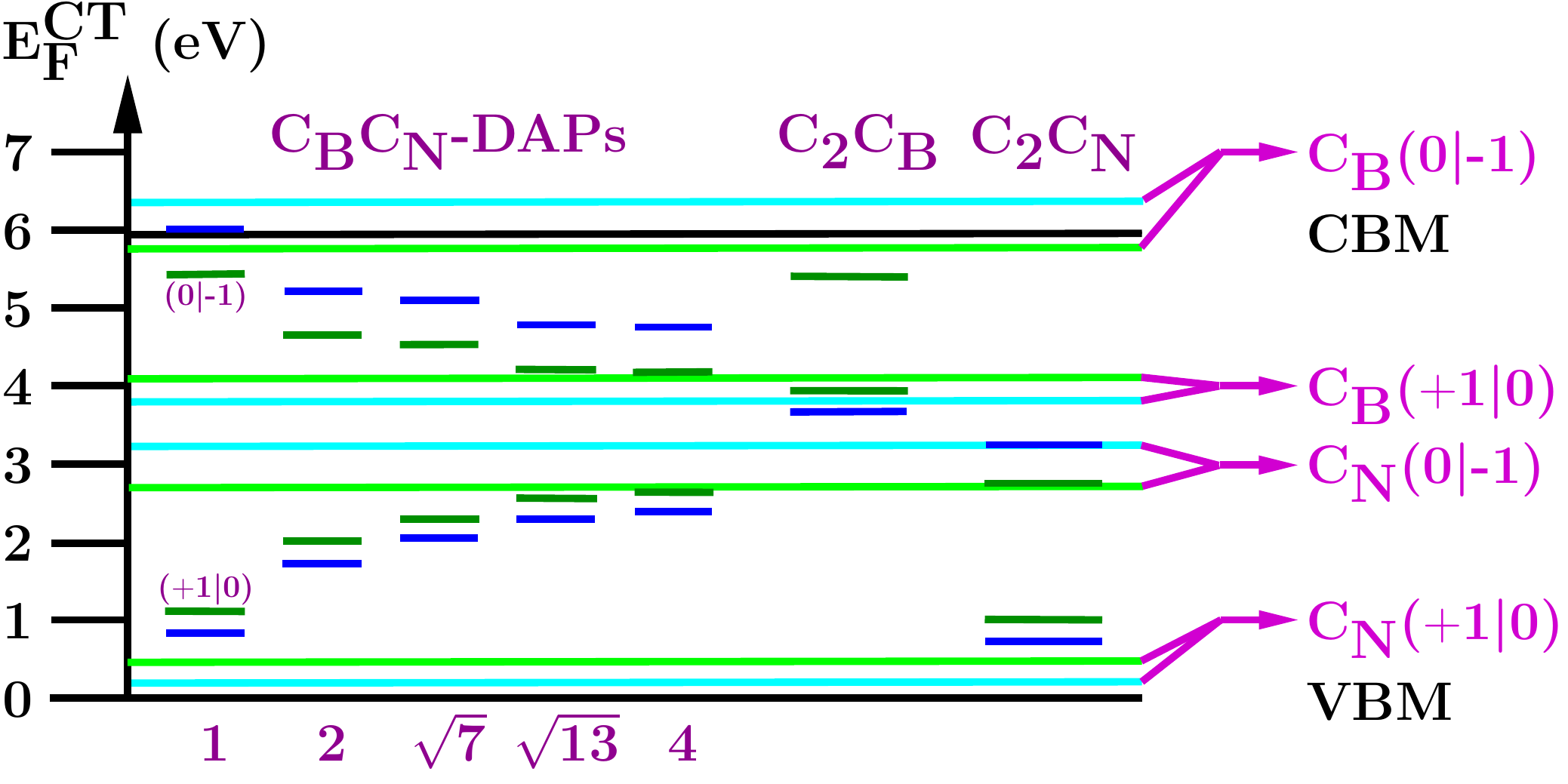}
  \caption{\small\label{Thermodynamic-CTLs-combined} Charge transition levels (CTLs) of substitutional carbon defects in bulk hBN
           relative to the valence band maximum (VBM) obtained directly from screened hybrid DFT (light green (point defects) and dark
           green (complexes)) and with \textit{a posteriori} charge corrections (light blue (point defects) and dark blue (complexes))
           \cite{PhysStatSol.B248(5).1067,PhysRevLett.102(1).016402(4)} (cf.\ text and Tab.~\ref{CTL-table})}
 \end{figure}

 Tab.~\ref{CTL-table} summarizes the CTL positions. All uncorrected CTLs lie within the band gap as displayed in 
 Fig.~\ref{Thermodynamic-CTLs-combined}. C$_\text{N}$(+1$\mid$0) resp.\ C$_\text{B}$(0$\mid$-1) is closest to the VBM resp.\ CBM. Since 
 the charge structure of the neutral DAPs is C$_\text{B}^+$-C$_\text{N}^-$ it is possible to either add an electron to C$_\text{B}^+$ 
 or to remove an electron from C$_\text{N}^-$ in the presence of the other charged point defect. Hence 
 C$_\text{B}$C$_\text{N}$(+1$\mid$0) resp.\ C$_\text{B}$C$_\text{N}$(0$\mid$-1) converges to C$_\text{N}$(0$\mid$-1) resp.\ 
 C$_\text{B}$(+1$\mid$0) with increasing separation. Meanwhile C$_\text{B}$C$_\text{N}$(+1$\mid$0) is raised whereas 
 C$_\text{B}$C$_\text{N}$(0$\mid$-1) is lowered. This is not only due to the diminishing, and finally vanishing, stabilizing effect of 
 the Coulomb interaction between the charged point defects on the neutral ground state (cf.\ Eq.~\eqref{CTL-calculation}). It is also 
 due to the decreasing energy difference between the orbitals involved in the charge transitions (highest occupied at C$_\text{N}^-$ 
 and lowest empty at C$_\text{B}^+$). As a consequence of this a ``forbidden region'' in which the Fermi level $E_\text{F}$ can be 
 modified without affecting the charge of the point defects and DAPs exists between C$_\text{N}$(0$\mid$-1) and 
 C$_\text{B}$(+1$\mid$0). The CTLs have also been calculated for the basic trimers C$_2$C$_\text{B}$ and C$_2$C$_\text{N}$. 
 C$_2$C$_\text{B}$(+1$\mid$0) and C$_2$C$_\text{N}$(0$\mid$-1) are similar to those of the isolated point defects without the 
 ``additional dimer'' C$_2$, whereas C$_2$C$_\text{B}$(0$\mid$-1) and C$_2$C$_\text{N}$(+1$\mid$0) resemble those of C$_2$.

 The charge corrections stabilize the charged systems by several hundreds of meV (cf.\ Tab.~\ref{CTL-table}) but the expected trend of 
 convergence is not obvious for the corrected CTLs (cf.\ Fig.~\ref{Thermodynamic-CTLs-combined}).

 
 Note that similar values for the corrected CTL positions of point defects and dimer were found by Mackoit-Sinkevi\v{c}ien\.{e} et al.\
 \cite{ApplPhysLett.115(21).212101(4)}, while Weston et al.\ \cite{PhysRev.B97(21).214104(13), PhysRev.B102(9).099903(2)} also give a 
 similar value for C$_\text{B}$(+1$\mid$0) but a somewhat lower one for C$_\text{N}$(0$\mid$-1). Just like in this work both employed a 
 bulk model, hybrid DFT and charge corrections according to Freysoldt et al.\ \cite{PhysStatSol.B248(5).1067, 
 PhysRevLett.102(1).016402(4)}.

 \begin{table}[h]
  \caption{\small\label{CTL-table} Charge transition levels (CTLs) of substitutional carbon defects in bulk hBN relative to the valence
  	       band maximum (VBM) obtained from screened hybrid DFT with (and without) \textit{a posteriori} charge corrections 
  	       \cite{PhysStatSol.B248(5).1067,PhysRevLett.102(1).016402(4)} (cf.\ text and Fig.~\ref{Thermodynamic-CTLs-combined})
  	       \footnote{\textcolor{ForestGreen}{green}: outside the indirect band gap (above the conduction band minimum (CBM)),
  	       \textcolor{blue}{blue}: close to the CBM, \textcolor{orange}{orange}: most likely above the CBM (too large oscillations in
  	       potential alignment)}}
  \begin{ruledtabular}
   \begin{tabular}{l|cc}
    Defect & CTL(+1$\mid$0) (eV) & CTL(0$\mid$-1) (eV) \\
    \hline
    C$_\text{B}$ & 3.81 (4.11) & \textcolor{ForestGreen}{6.39} (5.79) \\
    C$_\text{N}$ & 0.21 (0.47) & 3.27 (2.72) \\
    C$_\text{B}$C$_\text{N}$-DAPs & \multicolumn{1}{c}{} & \\
               1 & 0.85 (1.13) & \textcolor{blue}{5.99} (5.42) \\
               2 & 1.73 (2.01) & 5.25 (4.68) \\
      $\sqrt{7}$ & 2.02 (2.31) & 5.10 (4.53) \\
     $\sqrt{13}$ & 2.29 (2.57) & 4.81 (4.23) \\
               4 & 2.37 (2.65) & 4.77 (4.20) \\
    Trimers & \multicolumn{1}{c}{} & \\
    C$_2$C$_\text{B}$ & 3.66 (3.94) & \textcolor{orange}{unclear} (5.40) \\
    C$_2$C$_\text{N}$ & 0.71 (0.98) & 3.29 (2.72) \\
   \end{tabular}
  \end{ruledtabular}
 \end{table}

\subsection{\label{cw-EPR-ODMR} cw ODMR spectra} 
 The cw ODMR spectrum, in particular the line shape resp.\ broadening, serves as a further criterion for identifying experimentally 
 isolated emitters. It has been analyzed for all ground state $S=1/2$ systems (doublets), i.e.\ the neutral point defects and basic 
 trimers as well as the singly charged DAPs (see also \cite{SM}).

 The theoretical ODMR spectra are mostly obtained based on a simplified spin Hamiltonian $\hat{H} = 
 \hat{H}_\text{EZI}+\sum_{k}\hat{H}_\text{HFI}(k)$ taking into account the interaction of the unpaired electron with the external 
 magnetic field $\textbf{B}$ (electron Zeeman interaction $\hat{H}_\text{EZI}$) and the spins of the nuclei $k$ (hyperfine interaction 
 $\sum_{k}\hat{H}_\text{HFI}(k)$). In the chosen approximation they are determined by the external magnetic field $\textbf{B}$ 
 (assuming a gyromagnetic tensor $g = 2*\mathbb{1}_3$ for the electron), the isotopic composition and 
 the associated hyperfine matrices $A(k)$ from hybrid DFT. The hyperfine parameters are calculated for the most abundant nuclear spin 
 active isotopes (${}^{11}$B, ${}^{14}$N, ${}^{13}$C). Since nuclear spin active carbon is rare (${}^{13}$C of 1.1\%), it is assumed 
 that only boron (${}^{11}$B of 80.1\%) and nitrogen (${}^{14}$N of 99.6\%) contribute to the hyperfine splitting. For the neutral 
 point defects and singly charged DAPs up to second neighbors (d=$\sqrt{3}$) of the paramagnetic carbon atom carrying the unpaired 
 electron are taken into account. For the neutral basic trimers those of each carbon atom were chosen instead. The results and the 
 justification of the underlying assumptions are discussed in the following.

 We also considered a full spin Hamiltonian for test cases, e.g., C$_\text{B}$ defect, which reads as 
 \begin{equation} \label{spin-Hamiltonian} 
  \hat{H} = \hat{H}_\text{EZI}+\sum_{k}(\hat{H}_\text{HFI}+\hat{H}_\text{NZI}+\hat{H}_\text{NQI})(k) \text{,} 
 \end{equation}
 which includes $\hat{H}_\text{NZI}$ nuclear Zeeman-term and $\hat{H}_\text{NQI}$ nuclear quadrupole term too. The terms can be in 
 Eq.~\eqref{spin-Hamiltonian} expressed as  
 \begin{equation}
  \begin{gathered}
  \label{eqs:Hspterms}
   \hat{H}_\text{EZI} = \mu_\text{B}\mathbf{B}^\text{T}g\hat{\mathbf{S}} \\
   \approx \mu_\text{B}
   \begin{pmatrix} B_{x} & B_{y} & B_{z} \end{pmatrix}
   \begin{pmatrix} 2 & & \\ & 2 & \\ & & 2 \end{pmatrix}
   \begin{pmatrix} \hat{S}_{x} \\ \hat{S}_{y} \\ \hat{S}_{z} \end{pmatrix} \\
    = 2\mu_\text{B}\langle\mathbf{B}.\hat{\mathbf{S}}\rangle \\
   \hat{H}_\text{HFI}(k) = h\hat{\mathbf{S}}^\text{T}A(k)\hat{\mathbf{I}}_{k} \\  
   \hat{H}_\text{NZI}(k) = -\mu_{k}g_{k}\langle\mathbf{B}.\hat{\mathbf{I}}_{k}\rangle \\
   \hat{H}_\text{NQI}(k) = h\hat{\mathbf{I}}_{k}^\text{T}Q(k)\hat{\mathbf{I}}_{k} \text{,}
  \end{gathered}
 \end{equation}
 where $\mu_\text{B}$ and $\mu_k$ are the Bohr-magneton of the electron and nuclei $k$, $g_k$ and $Q(k)$ are the gyromagnetic factor 
 and quadrupole moment of nuclei $k$. In terms of broadening of the ESR signal, the most important parameter is the hyperfine tensor 
 $A(k)$ for nuclei $k$. We will show that $\hat{H}_\text{NZI}$ and $\hat{H}_\text{NQI}$ can be typically neglected at non-zero magnetic 
 fields, although, $\hat{H}_\text{NQI}$ can contribute to the final shape of the ESR or ODMR spectrum coming from the ${}^{14}$N nuclei 
 spins.

\begin{figure}[h]
  \includegraphics[width=\linewidth]{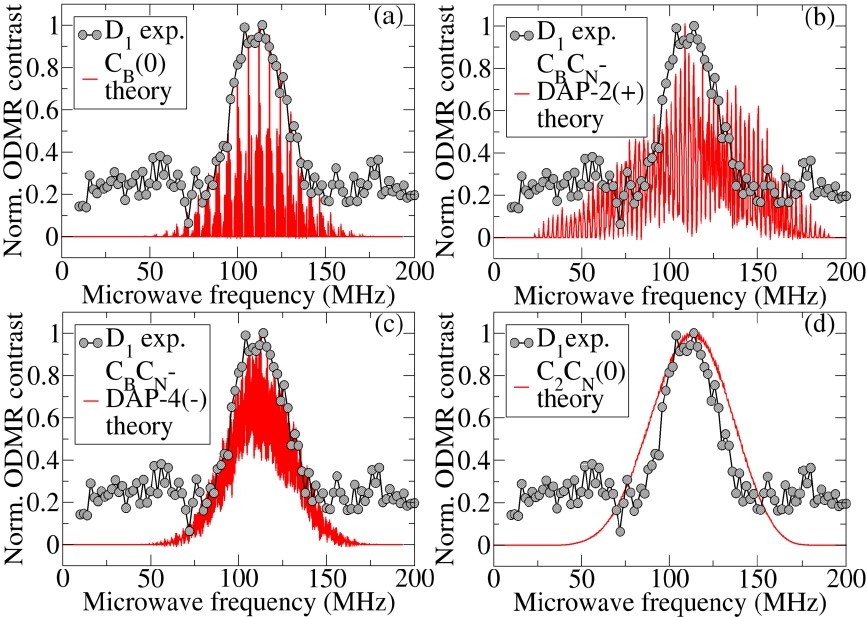}
  \caption{\small\label{D1-ODMR-compared-to-candidates} Measured ODMR spectrum of isolated optical emitter D$_1$ \cite{NatMat-1sthBN}
   	       compared to line broadening calculated for the defect identification candidates C$_\text{B}$(0),
   	       C$_\text{B}$C$_\text{N}$-DAP-2(+), C$_\text{B}$C$_\text{N}$-DAP-4(-), C$_2$C$_\text{N}$(0) in bulk hBN, all at an external
   	       magnetic field of 42~Gauss, theoretical spectra shifted to match the experimental peak position (cf.\ text)}
 \end{figure}

 Tab.~\ref{ODMR-line-broadening} summarizes the calculated line broadening and peak positions at an external magnetic field of 42~Gauss 
 as used for the ODMR measurement of D$_1$ spin center (cf.\ Fig.~\ref{D1-ODMR-compared-to-candidates} and 
 Ref.~\onlinecite{NatMat-1sthBN}). The line broadening FWHM$_\text{(Gauss)}$=2$\sqrt{2\ln(2)}\sigma_\nu$ is defined as the FWHM of a 
 Gaussian normal distribution with the same standard deviation $\sigma_\nu$. The original peak position due to the electron Zeeman 
 interaction is at 118~MHz (cf.\ Tab.~\ref{ODMR-theory-approaches}). The asymmetric influence of the hyperfine interaction induces a 
 defect-dependent blueshift of up to 15~MHz. Neutral C$_\text{N}$ exhibits a much larger line broadening (74~MHz) than neutral 
 C$_\text{B}$ (43~MHz) as expected in view of the hyperfine constants (cf.\ Tabs.~\ref{CB-hyperfine} and \ref{CN-hyperfine} and 
 \cite{SM}). The underlying reason is that the gyromagnetic ratio of the boron isotope ($\gamma_{I}/2\pi$ = 1366~Hz/Gauss for 
 ${}^{11}$B), which is a first neighbor of the paramagnetic carbon for C$_\text{N}$ but only a second neighbor for C$_\text{B}$, is 
 more than four times larger than that of the nitrogen isotope ($\gamma_{I}/2\pi$ = 308~Hz/Gauss for ${}^{14}$N). In view of the charge 
 structure the line broadening of the singly positive resp.\ negative DAP has to converge to that of neutral C$_\text{N}$ resp.\ 
 C$_\text{B}$ with increasing separation. While no unambiguous trend is found for the former, the presence of the charged point defect 
 increases it for the latter. The line broadening of the basic trimers is smaller than that of the corresponding point defect without 
 the adjacent dimer. We note that the C$_2$C$_\text{N}$ defect create a dense eigenvalue spectrum of the spin Hamiltonian with contribution of 14 nuclear spins which will result in a relatively smooth ODMR spectrum in the simulation.

 \begin{table}[h]
  \caption{\small\label{ODMR-line-broadening} Calculated ODMR line broadening FWHM$_\text{(Gauss)}$ (and peak center $\overline{\nu}$) 
           for $S=1/2$ substitutional carbon defects in bulk hBN at an external magnetic field of 42~Gauss:
   	       FWHM$_\text{(Gauss)}$=2$\sqrt{2\ln(2)}\sigma_\nu$ is the FWHM of a Gaussian normal distribution with the same standard 
   	       deviation $\sigma_\nu$. $\overline{\nu}$ is the expectancy value of the microwave frequency. See also \cite{SM}.}
  \begin{ruledtabular}
   \begin{tabular}{l|cc}
	Defect type & \multicolumn{2}{c}{FWHM$_\text{(Gauss)}$ (MHz) ($\overline{\nu}$ (MHz))} \\
	\hline
	Point defects & C$_\text{N}$(0) & C$_\text{B}$(0) \\
	& 74 (132) & 43 (119) \\
	C$_\text{B}$C$_\text{N}$-DAPs & positive (+) & negative (-) \\
	4           & 74 (132) & 42 (119) \\
	$\sqrt{13}$ & 73 (132) & 44 (119) \\
	$\sqrt{7}$  & 77 (133) & 49 (119) \\
	2           & 71 (131) & 54 (119) \\
	1 (dimer)   & 77 (125) & 70 (118) \\
	Trimers & C$_2$C$_\text{N}$(0) & C$_2$C$_\text{B}$(0) \\
	& 51 (127) & 27 (119) \\
   \end{tabular}
  \end{ruledtabular}
 \end{table}

 The experimental ODMR spectrum of D$_1$ is compared to those calculated for the defect identification candidates C$_\text{B}$(0),
 C$_\text{B}$C$_\text{N}$-DAP-2(+), C$_\text{B}$C$_\text{N}$-DAP-4(-), C$_2$C$_\text{N}$(0) (cf.\ 
 Fig.~\ref{D1-ODMR-compared-to-candidates}). It is concluded that D$_1$ has to be either C$_\text{B}$(0) or at least a very similar 
 structure where the spin density is localized around C$_\text{B}$. As previously discussed the theoretical model fulfills the 
 conditions found by Chejanovsky et al.\ \cite{NatMat-1sthBN} and all defect identification candidates exhibit PL in good agreement 
 with the measured spectrum. However C$_\text{B}$(0) is the simplest and best explanation of the observed ODMR, while non-similar 
 structures provide only inferior spectra.
 The observed ODMR peak is centered at 112-113~MHz. A constant shift of every single calculated spectrum has been introduced to match 
 this value, since the magnetic field measurement (42~Gauss) may have an error of a few Gauss. For C$_\text{B}$(0) and 
 C$_\text{B}$C$_\text{N}$-DAP-4($-$) ($-7$~MHz) it is smaller than for C$_2$C$_\text{N}$(0) ($-14$~MHz) and
 C$_\text{B}$C$_\text{N}$-DAP-2(+) ($-19$~MHz).
 While C$_\text{B}$(0) with 43~MHz line broadening yields a very good reproduction of the observed peak, the spectra of 
 C$_2$C$_\text{N}$(0) (51~MHz) and especially C$_\text{B}$C$_\text{N}$-DAP-2(+) (71~MHz) are clearly broader. In view of the charge 
 structure the results for C$_\text{B}$(0) and C$_\text{B}$C$_\text{N}$-DAP-4($-$) are close to each other. In both defects, the spin 
 density is localized around C$_\text{B}$, so one can conclude that C$_\text{B}$ defect should be involved in the D$_1$ spin center.

 Although the experimental background might reach up to 40\% of the peak maximum, it can not lead to another interpretation of the 
 results or a substantial reduction of their quality. The measured intensities within, and only within, the interval 92-134~MHz are 
 $>$40\% while they come close to this value at some other frequencies. It is hence an upper bound for the background. Since even after 
 subtraction of the full 40\% (as background correction) the normalized peak within the relevant interval is only slightly less broad, 
 so the conclusions remain unchanged.
 
 The hyperfine constants of C$_\text{B}$(0) are given in Tab.~\ref{CB-hyperfine}. Note that the involvement of nuclear spin active 
 carbon ($\gamma_{I}/2\pi$ = 1071~Hz/Gauss for ${}^{13}$C) would substantially split the ODMR peak of C$_\text{B}$(0) to a doublet 
 because of the huge spin density on the carbon atom attributed to the unpaired electron localized on its $p_z$ orbital. This behavior 
 is also found for the other $S=1/2$ defects \cite{SM}. Measuring it would provide further insight into the structure of D$_1$ and/or 
 other experimentally isolated emitters.

 The hyperfine parameters include the contribution of the spin polarization of the core electrons to the Fermi contact interaction.
 Tab.~\ref{CN-hyperfine} takes neutral C$_\text{N}$ as an example to show that this contribution can be significant, in particular if 
 the carbon is nuclear spin active. Without it Sajid et al.\ \cite{PhysRev.B97(6).064101(9)} found hyperfine constants in reasonable 
 agreement employing a 2D model and the regular HSE06 functional. Our calculations imply that the core polarization should be included 
 in the calculations of accurate hyperfine tensors.

 \begin{table}[h]
  \caption{\small\label{CB-hyperfine} Hyperfine constants ($A_{xx}$, $A_{yy}$, $A_{zz}$) of the neutral C$_\text{B}$ point defect in 
  	       bulk hBN up to second neighbors obtained from screened hybrid DFT (\# $\coloneqq$ numbers of atoms, d $\coloneqq$ distance
  	       within ideal lattice, cf.\ text and Fig.~\ref{CB-and-CN})}
  \begin{ruledtabular}
   \begin{tabular}{c|ccrrr}
 	total & \# & d(C$_\text{B}$) & $A_{xx}$ (MHz) & $A_{yy}$ (MHz) & $A_{zz}$ (MHz) \\
 	\hline
 	C & 1 & 0 & 12.1 & 12.1 & 231.3 \\ 
 	N & 3 & 1 & -9.0 & -5.1 & -9.0 \\ 
 	B & 6 & $\sqrt{3}$ & 1.4 & -0.9 & 6.1 \\ 
   \end{tabular}
  \end{ruledtabular}
 \end{table}


 \begin{table*}
  \caption{\small\label{CN-hyperfine} Hyperfine constants ($A_{xx}$, $A_{yy}$, $A_{zz}$) of the neutral C$_\text{N}$ point defect in 
           bulk hBN up to second neighbors obtained from screened hybrid DFT ($A_{1c}$ $\coloneqq$ core contribution, () $\coloneqq$ 
           value without $A_{1c}$, \# $\coloneqq$ numbers of atoms, d $\coloneqq$ distance within ideal lattice, cf.\ text and 
           Fig.~\ref{CB-and-CN})}
  \begin{ruledtabular}
   \begin{tabular}{c|ccrrrr}
    total & \# & d(C$_\text{N}$) & $A_{xx}$ (MHz) & $A_{yy}$ (MHz) & $A_{zz}$ (MHz) & $A_{1c}$ (MHz) \\
    \hline
    C & 1 & 0 & -19.2 (44.8) & -19.2 (44.8) & 156.5 (220.5) & -64.0 \\ 
    B & 3 & 1 & -17.9 (-16.7) & -16.0 (-14.8) & -24.5 (-23.3) & -1.2 \\ 
    N & 6 & $\sqrt{3}$ & -0.2 (0.8) & -0.8 (0.1) & 2.4 (3.4) & -1.0 \\ 
   \end{tabular}
  \end{ruledtabular}
 \end{table*}

 The influence of the isotopic composition on the ODMR spectra of the point defects (see Tabs.~\ref{CB-isotopic-composition} and 
 \ref{CN-isotopic-composition}) was analyzed, because ${}^{10}$B is also significantly abundant (19.9\%). Replacing ${}^{11}$B with
 ${}^{10}$B gradually reduces the line broadening, because the hyperfine matrices scale linearly with the nuclear gyromagnetic factor
 (g$_\text{n}$ = $0.600$ instead of $1.792$) although the nuclear spin is increased ($I = 3$ instead of $3/2$) 
 \cite{isotopes-replacement-order}. This effect is more pronounced for neutral C$_\text{N}$, since the boron isotopes are first 
 neighbors of the paramagnetic carbon. In this case it also affects the peak center which is slightly redshifted. Considering only 
 ${}^{11}$B is usually a good approximation, since large changes of the ODMR spectra (dominant contribution of ${}^{10}$B) are 
 unlikely to be observed at the natural distribution of isotopes.

 \begin{table}[h]
  \caption{\small\label{CB-isotopic-composition} Calculated effect of the isotopic composition on ODMR line broadening FWHM (and peak 
  	       center $\overline{\nu}$) for the neutral C$_\text{B}$ point defect in bulk hBN at an external magnetic field of 42~Gauss
  	       (boron isotopes are 2nd neighbors (d = $\sqrt{3}$) of C$_\text{B}$, p $\coloneqq$ abundance of isotopic composition)}
  \begin{ruledtabular}
   \begin{tabular}{c|crrr}
	\#(${}^{11}$B) & \#(${}^{10}$B) & p(\%) & $\overline{\nu}$ (MHz) & FWHM (MHz) \\
	\hline
	6 & 0 & 26.41 & 119 & 43 \\ 
	5 & 1 & 39.37 & 119 & 41 \\ 
	4 & 2 & 24.45 & 119 & 39 \\
    3 & 3 & 8.10 & 119 & 37 \\
	2 & 4 & 1.51 & 119 & 34 \\
	1 & 5 & 0.15 & 119 & 32 \\ 
	0 & 6 & 0.01 & 119 & 29 \\
   \end{tabular}
  \end{ruledtabular}
 \end{table}

 \begin{table}[h]
  \caption{\small\label{CN-isotopic-composition} Calculated effect of the isotopic composition on ODMR line broadening FWHM (and peak 
  	       center $\overline{\nu}$) for the neutral C$_\text{N}$ point defect in bulk hBN at an external magnetic field of 42~Gauss 
  	       (boron isotopes are 1st neighbours ($d = 1$) of C$_\text{N}$, p $\coloneqq$ abundance of isotopic composition)}
  \begin{ruledtabular}
   \begin{tabular}{c|crrr}
   	\#(${}^{11}$B) & \#(${}^{10}$B) & p (\%) & $\overline{\nu}$ (MHz) & FWHM (MHz) \\
	\hline
	3 & 0 & 51.39 & 132 & 74 \\
	2 & 1 & 38.30 & 129 & 66 \\
	1 & 2 & 9.52 & 126 & 57 \\
	0 & 3 & 0.79 & 123 & 45 \\
   \end{tabular}
  \end{ruledtabular}
 \end{table}

\begin{figure}[h]
  \includegraphics[width=\linewidth]{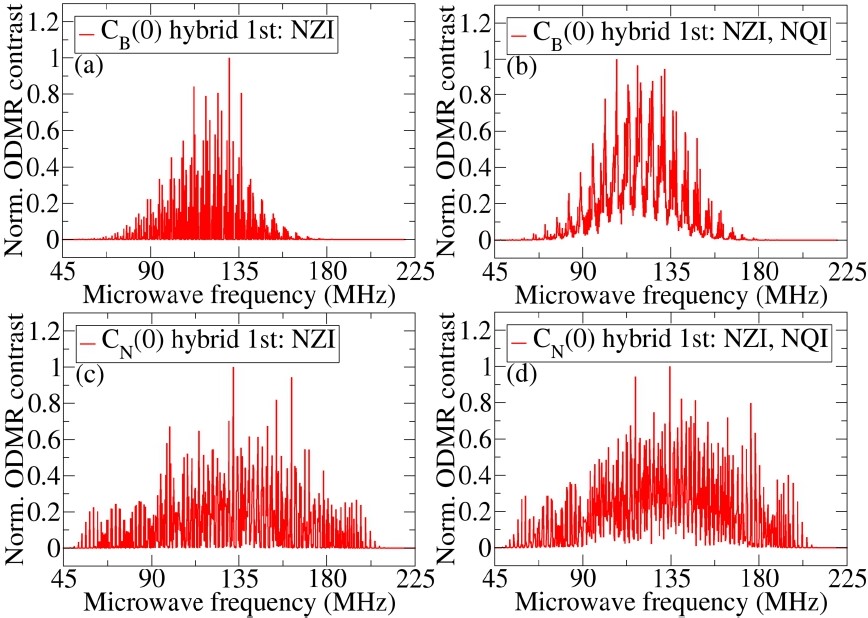}
  \caption{\small\label{point-defects-NQI-1st} Influence of the nuclear quadrupole interaction (NQI) on the theoretical cw EPR/ODMR 
   	       spectrum of the neutral point defects in bulk hBN at an external magnetic field of 42~Gauss (a/c) C$_\text{B}$/C$_\text{N}$
   	       without NQI (b/d) C$_\text{B}$/C$_\text{N}$ with NQI for 1st neighbours (see text for explanation, cf.\ Tab.\ 
   	       \ref{ODMR-theory-approaches})}
 \end{figure}

 The peak broadening is essentially determined by the hyperfine interaction and the simplified spin Hamiltonian (evaluated with
 second-order perturbation theory) yields already a good approximation of the ODMR spectra. This is demonstrated using the most basic 
 example of the neutral point defects C$_\text{B}$ and C$_\text{N}$. However it has to be kept is mind that a comprehensive discussion 
 of the full spin Hamiltonian is beyond the scope of the present study. Tab.~\ref{ODMR-theory-approaches} summarizes the results for 
 line broadening and peak center position obtained from different approaches. Further simplifying the spin Hamiltonian via using only 
 the hyperfine constants, instead of the full hyperfine tensor $A$ is not justified. It leads to a significant overestimation of the 
 line broadening, in particular for C$_\text{N}$ (38~MHz) whose peak center is additionally redshifted (5~MHz). Since second-order 
 perturbation theory cannot be combined with exact diagonalization for specific nuclei in the code, first-order perturbation theory is 
 also considered. Since the C$_\text{N}$ peak center is redshifted by 14~MHz, this seems to be a non-accurate approach. Exact 
 diagonalization allows to include the nuclear Zeeman interaction and additionally the nuclear quadrupole interaction, which have an 
 indirect effect on the electronic transitions as the HFI couples the electronic spin to the nuclear spins. Due to computational 
 expense which scales exponentially with the number of spins in the Hilbert space it was performed only for the first (1st) neighbors 
 of the respective point defect. Including the NZI reverses the 14~MHz redshift of the C$_\text{N}$ peak center. However this might 
 also be related to the underlying change of the algorithm to solve the spin Hamiltonian. Note that the NZI leads to core transition 
 peaks in a low frequency range ($<$30MHz) that was not included in the calculation of peak center and line broadening. Judging from 
 Fig.~\ref{point-defects-NQI-1st} the NQI can still have a non-negligible effect on the spectral shape although peak center and line 
 broadening are hardly affected (cf.\ Tab.~\ref{ODMR-theory-approaches}). Indeed the entries of the $Q$ matrices (cf.\ 
 Eqs.~\eqref{V-to-Q} and \eqref{spin-Hamiltonian} and \cite{SM}) are $\le$1MHz. Hence the line splittings induced by the NQI can only 
 be correspondingly small. However it can also induce significant intensity redistribution explaining its importance in the given 
 context. The results for C$_\text{B}$ are further evidence for its near-identification as experimentally isolated optical emitter 
 D$_1$.

 \begin{table}[h]
  \caption{\small\label{ODMR-theory-approaches} ODMR line broadening FWHM (and peak center $\overline{\nu}$) for the neutral point 
   	       defects (C$_\text{B}$, C$_\text{N}$) in bulk hBN at an external magnetic field of 42~Gauss calculated with different 
   	       approaches (see text for explanation). Labels ``perturb2'' and ``perturb1'' are second-order and first-order perturbation 
   	       theories whereas ``hybrid'' means an exact diagonalization for NZI and NQI terms but the other terms are considered within 
   	       second-order perturbation theory.}
  \begin{ruledtabular}
   \begin{tabular}{l|cc}
	Approach & \multicolumn{2}{c}{FWHM (MHz) ($\overline{\nu}$ (MHz))} \\
	\hline
	                         & C$_\text{N}$(0)      & C$_\text{B}$(0) \\
	EZI only                 & \,\,\,\, 0 (118)     & \, 0 (118) \\
    perturb2 ($A$ constants) & \, 112 (127) \,      & 49 (119) \\
	perturb2 (full $A$)      & \,\,\,\, 74 (132) \, & 43 (119) \\
	perturb1 (full $A$)      & \,\,\,\, 73 (118) \, & 43 (118) \\
    hybrid (1st: NZI)        & \,\,\,\, 72 (132) \, & 42 (119) \\
	hybrid (1st: NZI, NQI)   & \,\,\,\, 72 (132) \, & 42 (119) \\
   \end{tabular}
  \end{ruledtabular}
 \end{table}

\subsection{\label{summary} Summary} 
 Paramagnetic substitutional carbon defects in hexagonal boron nitride have been characterized via modeling their photoluminescence 
 spectra, charge transition levels and cw ODMR spectra. Clear trends in all of these were found by analyzing the separation between 
 C$_\text{B}$ and C$_\text{N}$. Furthermore a near-identification of an experimentally isolated optical emitter D$_1$ as neutral 
 C$_\text{B}$ is given.

 $E_\text{ZPL}$=1.695~eV (2.468~eV) was obtained for neutral C$_\text{B}$ (C$_\text{N}$) which are approximate values due to the finite 
 size effects and the accuracy in the applied DFT functional. $E_\text{ZPL}$ of the singly negative (positive)
 C$_\text{B}$C$_\text{N}$-DAP converges to this, or at least a similar, value from below with increasing separation. This behavior is
 explained based on the relevant electronic transition and the charge structure. The result for the neutral dimer 
 (C$_\text{B}$C$_\text{N}$, $E_\text{ZPL}$=4.131~eV), which allegedly causes the ubiquitous 4.1-eV photoluminescence 
 \cite{ApplPhysLett.115(21).212101(4)}, is consistent with previous hybrid DFT results \cite{ApplPhysLett.115(21).212101(4),
 JPhysChem.A125(6).1325}. $E_\text{ZPL}$ of the neutral C$_\text{B}$C$_\text{N}$-DAPs has to decrease with separation due to the 
 position of the states inside the band gap (see also \cite{PhysRev.B103(11).115421(13)}). Each basic trimer has a lower $E_\text{ZPL}$ 
 than its ``building blocks'' point defect and dimer as expected in view of the overlap of their states inside the band gap. 

 An intense feature in PSB next to ZPL at 0.15-0.17~eV (except for the neutral dimer at 0.20~eV) was found for all defects. Similar 
 results were obtained in theoretical \cite{JPhysChem.A125(6).1325,PhysRev.B103(11).115421(13)} as well as experimental 
 \cite{NatMat-1sthBN,NatMat.20(3).321,PhysRev.B78(15).155204(8),NanoLett.16(7).4317} studies. In particular ODMR experiments yielded 
 evidences for the connection of these features to carbon-related defects \cite{NatMat.20(3).321, NatMat-1sthBN}. Four candidates for 
 the identification of D$_1$ spin center were provided based on the agreement with the ZPL energies within 0.2~eV accuracy.

 The cw ODMR spectra for all ground state $S=1/2$ systems have been calculated at 42~Gauss external magnetic field based on a 
 simplified spin Hamiltonian. The asymmetric influence of the hyperfine interaction induces a defect-dependent blueshift of the peak 
 positions from 118~MHz up to 133~MHz. C$_\text{N}$ exhibits a much larger line broadening than C$_\text{B}$ due to the hyperfine 
 parameters. Replacing ${}^{11}$B with the less abundant ${}^{10}$B reduces the peak width and can also shift the peak center. However 
 large changes are unlikely to be observed at the natural distribution of isotopes. The charge structure explains the convergence of 
 the line broadening of the singly charged C$_\text{B}$C$_\text{N}$-DAPs.

 The calculated ZPL energy and PSB as well as the ESR spectrum strongly implies that C$_\text{B}$ is involved as spin-active part of 
 the D$_1$ spin center, and it could be the isolated neutral C$_\text{B}$ defect. Measuring the ODMR splitting due to ${}^{13}$C 
 isotope would provide further insight about the chemical composition of D$_1$ or other emitters.

 An exemplary analysis of the point defects demonstrates that the ODMR line broadening is essentially due to hyperfine interaction. The
 simplified spin Hamiltonian together with second-order perturbation theory results in a good approximation of the spectra. Using only 
 the hyperfine constants, i.e.\ neglecting the Euler angles, is not appropriate as it leads to a large overestimation of the line 
 broadening. A hybrid approach allowed to include the nuclear Zeeman and additionally the quadrupole interaction for the first 
 neighbors. The line broadening remained essentially unaffected. Nuclear quadrupole interaction is important for an exact description
 of the spectral shape as it can lead to non-negligible modifications of the intensities but it does not effect the overall broadening 
 of the ESR signal.

\section{\label{final} Conclusions and outlook} 
 The modeling of PL and ODMR spectra has yielded helpful criteria for the identification of paramagnetic substitutional carbon defects 
 in hBN. An example for a near-identification as neutral C$_\text{B}$ has been given. The separation between C$_\text{B}$ and 
 C$_\text{N}$ emerged as key to control the ZPL wavelength and charge compensation of defects. All findings are further steps towards 
 the identification of these structures and later their target-oriented engineering as quantum bits. 

 It is demonstrated that first principles theory is able to predict the complex electron spin resonance spectrum of defects in hBN. It 
 has been found that the full hyperfine tensor of the nuclei is required to accurately determine the linewidth and the central position 
 of the ODMR signal at a given external magnetic fields. As direct measurement of the full hyperfine tensors is challenging our 
 results demonstrate that the accurate interpretation of ESR and ODMR signals in hBN requires a tight cooperation between experimental 
 and \textit{ab initio} theoretical spectroscopy.

\section*{Acknowledgements}
 We are grateful to Prof.~J\"org Wrachtrup and Dr.~Durga Dasari for sharing and explaining their experimental results, Christoph 
 Freysoldt for discussions about his charge correction scheme and EasySpin creator Stefan Stoll for helping us to use his program. This 
 research was supported by the Ministry of Innovation and Technology and the National Research, Development and Innovation Office 
 (NKFIH) within the Quantum Information National Laboratory of Hungary, the National Quantum Technology Program (NKFIH Grant 
 No.~2017-1.2.1-NKP-2017-00001), the National Excellence Program (NKFIH Grant No.~KKP129866), as well as the support of European 
 Commission within the Quantum Technology Flagship Project ASTERIQS (Grant No.~820394).  We acknowledge that the results of this 
 research have been achieved using the DECI resource Eagle HPC based in Poland at Poznan with support from the PRACE aisbl and 
 resources provided by the Hungarian Governmental Information Technology Development Agency, including the project ``gallium''.
 


\begin{thebibliography}{42}
\bibitem [{iso()}]{isotopes-replacement-order}%
  \BibitemOpen
  \href@noop {} {}\bibinfo {note} {At this point the order of replacement is
  irrelevant in view of the nuclear symmetry and the neglect of core-core
  interaction in the spin Hamiltonian.}\BibitemShut {Stop}%
\end{thebibliography}

%

\end{document}